\documentclass[aps,showpacs]{revtex4}
\usepackage{psfig}

\begin{document}
%preprint number:
%
\title{
\[ \vspace{-2cm} \]
\noindent\hfill\hbox to 1.5in{\rm  } \vskip 1pt
\noindent\hfill\hbox to 1.5in{\rm SLAC-PUB-10269\hfill  } \vskip 1pt
\noindent\hfill\hbox to 1.5in{\rm December 19, 2003 \hfill}\vskip 10pt
Quantized Cosmology II: de Sitter Space\footnote{This work was supported by the
U.~S.~DOE, Contract No.~DE-AC03-76SF00515.}}
\author{Marvin Weinstein and Ratin Akhoury\footnote{On sabbatical leave from
Dept. of Physics, University of Michigan, Ann Arbor, MI 48109-1120}}
\address{Stanford Linear Accelerator Center, Stanford University,
  Stanford, California 94309}
\date{December 19, 2003}
\begin{abstract}
This work applies the formalism developed in our earlier paper\cite{Weinstein:2003ya}
to de Sitter space.  After exactly solving the relevant Heisenberg equations
of motion we give a detailed discussion of the subtleties associated with defining
physical states and the emergence of the classical theory.  This computation
provides the striking result that quantum corrections to this long wavelength limit
of gravity eliminate the problem of the {\it big crunch\/}.  We also show that
the same corrections lead to possibly measureable effects on the CMB radiation.
Finally, for the sake of completeness we discuss the special case, $\Lambda=0$, and
its relation to Minkowski space.
\end{abstract}
% slac pub 10244
\maketitle

\newcommand{\ba}{\begin{eqnarray}}
\newcommand{\ea}{\end{eqnarray}}
\newcommand{\x}{\mbox{$\vec{x}$}}
\newcommand{\dphidt}{{\epsilon d\phi(t,\x) \over dt}}
\newcommand{\Phidot}{{d\Phi(t) \over dt}}
\newcommand{\Phiddot}{{d^2\Phi(t) \over dt^2}}
\newcommand{\adot}{{da(t) \over dt}}
\newcommand{\addot}{{d^2a(t) \over dt^2}}
\newcommand{\gradphi}{\epsilon\vec{\nabla}\phi(t,\x)}
\newcommand{\gradphisq}{\epsilon^2\vec{\nabla}\phi(t,\x)\cdot\vec{\nabla}\phi(t,\x)}
\newcommand{\be}{\begin{equation}}
\newcommand{\ee}{\end{equation}}
\newcommand{\goo}{(1 + 2\epsilon\,\chi(t,\vec{x}))}
\newcommand{\gxx}{(1 - 2\epsilon\,\chi(t,\vec{x}))}
\newcommand{\gooinv}{{1\over 1 + 2\epsilon\,\chi(t,\vec{x})}}
\newcommand{\gxxinv}{a(t)^2\,({1 - 2\epsilon\,\chi(t,\vec{x})})}
\newcommand{\udot}{{d u(t)\over dt}}
\newcommand{\uddot}{{d^2u(t) \over dt^2}}
\newcommand{\Hub}{{\cal H}}
\newcommand{\Hdot}{{d{\cal H}(t) \over dt}}
\newcommand{\Hddot}{{d^2{\cal H}(t) \over dt^2}}
\def\ket#1{\vert #1 \rangle}
\def\bra#1{\langle #1 \vert}
\def\vev#1{\left< #1 \right>}
\def\A{\frac{3\kappa^2}{16{\bf V}}}

%\narrowtext
\section{Introduction}

In a recent paper\cite{Weinstein:2003ya} we presented a quantum
mechanical formalism for the part of the computation of the anisotropy in the
CMB radiation\cite{Mukhanov:PRpt215} that is usually treated purely classically.
In this paper we apply this formalism to the case of
de Sitter space.  There are two reasons for doing this:  first,
the problem is interesting in its own right; second, it is exactly
solvable and the solution clarifies subtle features of the discussion
given in our first paper.

Our earlier discussion began by assuming the usual Friedmann-Robertson-Walker
metric in homogeneous isotropic coordinates,
\be
   ds^2 = - dt^2 + a(t)^2 d\vec{x}\cdot d\vec{x} ,
\label{FRW}
\ee
and an action of the form
\be
{\cal S} = {\bf V} \left[ - {3 \over \kappa^2} a(t) \left({d a(t) \over dt}\right)^2
+ {1\over 2} a(t)^3 \left({ d\Phi(t) \over dt} \right)^2 - a(t)^3 V(\Phi(t)) \right] .
\label{FRWact}
\ee
We then introduced the change of variables $u(t)^2=a(t)^3$ and rewrote the action
in the simpler form
\be
{\cal S} = {\bf V} \left[ - {4 \over 3\kappa^2} \left({d u(t) \over dt}\right)^2
+ {1\over 2} u(t)^2 \left({ d\Phi(t) \over dt} \right)^2 - u(t)^2 V(\Phi(t)) \right].
\ee
The rest of our discussion followed from canonically quantizing this theory and
seeing how much of the Einstein equations could be recovered at the level of the
Heisenberg equations of motion.

This paper follows the same steps, but for an action in which
$V(\Phi)$ is replaced by a cosmological constant $\Lambda$; i.e.,
\be
{\cal S} = {\bf V} \left[ - {4 \over 3\kappa^2} \left({d u(t) \over dt}\right)^2
 - u(t)^2 \Lambda \right].
\ee

\section{Solving The Heisenberg Equations of Motion}

Direct commutation of the Hamiltonian with the operators $u(t)$ and $p_u(t)$ yields the
Heisenberg equations of motion
\be
{\bf H} = - {3 \kappa^2 \over 16 {\bf V} } p_u^2 +  {\bf V} u^2 \Lambda
\ee
and the Hamilton equations of motion for $u(t)$ and $p_u(t)$ are
\be
    {du(t)\over dt} = -\frac{3\kappa^2}{8{\bf V}} p_u \,;\,
    {d^2u(t) \over dt^2} = \frac{3\kappa^2 \Lambda}{4} u .
\ee
The exact solutions to these equations, written in terms of the operators $u(t=0)=u$ and $p_u(t=0)=p_u$
are
\ba
u(t) &=& \cosh(\omega t) u - \frac{3\kappa^2}{8 {\bf V} \omega} \sinh(\omega t) p_u
\nonumber\\
p_u(t) &=& \cosh(\omega t) p_u -\frac{8 {\bf V} \omega}{3\kappa^2}\sinh(\omega t) u ,
\label{exsols}
\ea
where we have defined
\be
    \omega = \sqrt{3\kappa^2 \Lambda \over 4} .
\ee

It is convenient to rewrite Eq.\ref{exsols} in terms of exponentials; i.e.,
\be
u(t) = \frac{e^{\omega t}}{2}\left( u - \frac{3\kappa^2}{8 {\bf V} \omega} p_u \right)+
\frac{3\kappa^2 e^{-\omega t}}{16 {\bf V} \omega}\,
\left( p_u + \frac{8 {\bf V}\omega}{3\kappa^2} u \right)
\ee
and to introduce the canonically conjugate asymptotic operators
\be
 u_\infty = \frac{1}{\sqrt{2}} \left( u-\frac{3\kappa^2}{8{\bf V}\omega}p_u
   \right) \,;\,
p_\infty = \frac{1}{\sqrt{2}}\left( p_u + \frac{8{\bf V} \omega}{3\kappa^2} u \right) .
\ee
In terms of these operators the solution for the operator $u(t)$ and the Hamiltonian take
the simple forms
\be
u(t) = \frac{1}{\sqrt{2}}\,e^{\omega t} u_\infty + \frac{1}{\sqrt{2}}\,
\frac{3\kappa^2}{8 {\bf V} \omega}\, e^{-\omega t} p_\infty ,
\label{exactu}
\ee
and
\be
{\bf H} = \frac{\sqrt{3\Lambda}\kappa}{4}\left( u_\infty p_\infty
+  p_\infty u_\infty \right) .
\label{hamiltonian}
\ee
From this point on all of the technical work is finished, the only chore which remains
is to extract the physical significance of these results.

\section{The Missing Friedmann Equation: Defining Physical States}

Before discussing the content of this solution, we must spend a
few moments defining the space of physical states.  This question
comes up because, as we pointed out in our previous paper, as a
consequence of working in a fixed coordinate system, we
don't obtain all of the Einstein equations as Heisenberg equations
of motion.  We showed that in the classical theory, the
missing equations could be imposed as constraints; since, as a consequence
of the equations of motion which we do have, it is
possible to prove that if they are satisfied at any one time, then
they are always satisfied.  Next, we showed that in the quantum theory
we could parallel the classical discussion and define a one parameter family of
operators, ${\bf G}_\alpha$, each of which satisfies an equation
of the form
\be
{3 u\over 4} \left({1\over {\bf A_\alpha}} {d {\bf
G}_\alpha\over dt} + 3 {\bf G}_\alpha\right) =0 .
\label{gaugeinv}
\ee
for some non-vanishing operator ${\bf A}_\alpha$ (where $0\le
\alpha \le 1$). Finally, we argued that, in contrast to the
classical situation, it doesn't make any sense to define the space
of physical states by the strong condition $G_\alpha(t)\ket{\Psi}=0$, for some
value of $\alpha$.
Instead, we stated that the correct condition is that $\ket{\Psi}$
is physical if and only if
\be
    \lim_{u(t)^2\rightarrow\infty} {\bf G}(t)\ket{\Psi} =0.
\label{ginv}
\ee
The nice thing about the example with just a cosmological constant is that we can easily
understand why we say that Eq.\ref{ginv} is the best one can do.

Adopting the same definition of the Hubble operator obtained in our earlier paper
\be
    {\cal H} = -{\kappa^2 \over 8{\bf V}}\left( p_u {1\over u} + {1\over u^3} p_u
     u(t)^2 \right)
\ee
and defining the operator ${\bf Q}$ to be
\be
{\bf Q}=-{3\kappa^4\over 64 {\bf V}^2 u^4} ,
\ee
it is a straightforward exercise in taking commutators to show that the one parameter
family of operators ${\bf G}_\alpha $ can be written as
\ba
    {\bf G}_\alpha &=& {\cal H}^2+ \alpha {\bf Q} - \frac{\kappa^2}{3}\Lambda \nonumber\\
    &=& \frac{\kappa^4}{16 {\bf V}^2 u^2} p_u^2 + (1-\alpha){3\kappa^4\over 64 {\bf V}^2 u^4}
    - \frac{\kappa^2}{3}\Lambda  \nonumber\\
    &=& -\frac{\kappa^2}{3 {\bf V} u^2}{\bf H} + (1-\alpha){3\kappa^4\over 64 {\bf V}^2 u^4} .
\label{galpha}
\ea
Noting that the Hamiltonian, ${\bf H}$, is time independent we have
\be
    {\bf G}_\alpha(t) = -\frac{\kappa^2}{3{\bf V}u(t)^2} \,\left[{\bf H}
    - (1-\alpha){9\kappa^2\over 64 {\bf V} u(t)^2} \right] .
\ee
Thus, we see that for the case $\alpha=1$, defining the space of physical states by
the condition ${\bf G}_1 \ket{\Psi} = 0$ is equivalent to
the Wheeler-Dewitt equation; i.e., ${\bf H}\ket{\Psi}=0$.
Unfortunately, the statement that the Hamiltonian is zero on this subspace
of states means they don't evolve.  However, this is in direct conflict
with the Heisenberg equations of motion (Eq.\ref{exsols}),
which is, of course, unacceptable.

On the other hand, if we choose another value of $\alpha$, to avoid
an immediate contradiction, we still run into trouble.   This is because
we can explicitly solve for such states by
using the explicit form of ${\bf G}_\alpha$ in Eq.\ref{galpha},
to rewrite the ${\bf G}_\alpha \Psi(u)=0$ as a differential equation
in $u$.  The result of this computation is that
the equation has no square integrable
solutions.  It therefore follows that there are no satisfactory
candidates for physical states which satisfy this strong form of the
constraint.  In contrast, given the exact solution for
$u(t)$, we see that any state for which ${\bf H}\ket{\Psi}$ has a
finite norm will, for sufficiently large $|t|$, satisfy Eq.\ref{ginv} to
arbitrary accuracy.  More precisely, any state $\ket{\Psi}$ such that
\be
    \bra{\Psi} {\bf H}^2 \ket{\Psi} < \infty
\label{finitee}
\ee
will satisfy the asymptotic condition
\be
    \lim_{t\rightarrow \pm\infty} {\bf G}(t) \ket{\Psi} = 0
\ee
This means that any Gaussian wave packet in $u_\infty$ will be a physical state.
It also means that for large times all the physics measured in such a state
will be compatible with the full set of Einstein equations.
In the next section we explicitly demonstrate this fact.

\section{Seeing the Classical Theory Emerge}

Now that we have defined the space of physical states, we turn
to a discussion of the only two physical observables in this
theory; the expansion rate and the volume of the universe. In what
follows we call an allowed quantum state a {\it quantum observer\/}.
What we wish to ascertain is to what degree the value of each of these
observables depends upon the {\it quantum observer\/}. Obviously, the
exact solution given in Eq.\ref{exactu} shows that at large times
the expansion rate is attached to the scale factor and is totally
independent of the observer.  This, however, is not true of the volume.
Thus, in the remainder of this section we will discuss the degree to
which the measured properties of the volume operator differ from
quantum observer to quantum observer.

Since we started off quantizing in a volume with coordinate size
${\bf V}$, the volume of the universe at any time is given by
\ba
    V(t) &=& {\bf V} u(t)^2 \cr
    &=& \frac{{\bf V}}{2}\, \left[
    e^{2\omega t} u_\infty^2
    + \left( \frac{3\kappa^2}{8{\bf V} \omega} \right)^2 \,e^{-2\omega t}
    p_\infty^2
    + \frac{3\kappa^2}{8{\bf V} \omega} \left(u_\infty p_\infty
    + p_\infty u_\infty\right)\right] . \label{voft}
\ea
A surprising feature of this formula is that for large
times in the past and future the volume operator $V(t)$ behaves
classically.  By the phrase {\it $V(t)$ behaves
classically\/}, we mean that if one measures $V(t)$ at some early
or late time $t_1$ and obtain a definite value, then we will be
able to predict the value we will obtain if we measure $V(t)$ at
some later time $t_2$.  A cursory examination of
Eq.\ref{voft} shows that for very large positive times $V(t)$ is,
to arbitrarily high accuracy, proportional to the single operator
$u_\infty^2$ (at large negative times it is proportional to
$p_\infty^2$).  Thus, for example, we see that a measurement of $V(t_1)$, for
sufficiently large $t_1$, corresponds to a measurement of $u_\infty^2$,
which means that we know $V(t)$ for all times $t_2 > t_1$.

From the fact that $u_\infty$ and $p_\infty$ are canonically conjugate
variables we see that if we were to try and identify a quantum observer
with an eigenstate of $p_\infty$, then the volume operator would be
well-determined in the past, but completely undetermined in the future.  Conversely,
eigenstates of $u_\infty$ correspond to states for which the volume operator
is completely well determined in the future, but completely undetermined in
the past.  Fortunately, the condition that physical states must be normalizeable
states for which Eq.\ref{finitee} holds, tells us that we cannot identify
such states with quantum observers.  States which can be identified with
quantum observers are Gaussian packets,
\be
\ket{\Psi} = e^{-\frac{\gamma}{2} u_\infty^2}
\ee
and the coherent states, $\ket{u_0,p_0,\gamma}$, obtained from them.
These coherent states are defined by
\be
    \ket{u_0,p_0,\gamma} = e^{i p_0 u_\infty}\,e^{-i u_0 p_\infty} \ket{\Psi},
\ee
and the expectation values of $u_\infty$ and $p_\infty$ in these states are given by
\be
\bra{u_0,p_0,\gamma} u_\infty \ket{u_0,p_0,\gamma} = u_0 ,\,
\bra{u_0,p_0,\gamma} p_\infty \ket{u_0,p_0,\gamma} = p_0.
\ee
Moreover, the relevant products of these operators have the values
\ba
\bra{u_0,p_0,\gamma} u_\infty^2 \ket{u_0,p_0,\gamma} &=& u_0^2 + \frac{1}{2\gamma},
\nonumber\cr
\bra{u_0,p_0,\gamma} p_\infty^2 \ket{u_0,p_0,\gamma} &=& p_0^2 + \frac{\gamma}{2}, \nonumber\cr
\bra{u_0,p_0,\gamma} u_\infty p_\infty + p_\infty u_\infty\ket{u_0,p_0,\gamma}
&=& 2\Re(\vev{u_\infty\,p_\infty}) = 2\,u_0 p_0  .
\ea
The nice thing about such coherent states is that
they are the kind of states we would expect to obtain if, in the past,
we make a measurement which determines $V(-t)$ to have a central value
$\frac{{\bf V}}{2}\,e^{\omega|t|} p_0^2$, with a width parameterized by $\gamma$.
For this same packet, measurements of $V(t)$ in the distant future will produce
results centered about the value $\frac{{\bf V}}{2}\,e^{\omega|t|} u_0^2$,
with a width parameterized by $1/\gamma$.

\section{Equivalence Classes of Observers}

From this point on we will restrict the term quantum observer to mean a coherent state
of the form defined above.  What we wish to discuss next is the fact that
many of these observers are equivalent to one another in a way which we will make
precise.  Begin by considering
\be
    \vev{V(t)} = \bra{u_0,p_0,\gamma} V(t) \ket{u_0,p_0,\gamma}
    = \frac{{\bf V}}{2} \left[ e^{2\omega t} \vev{u_\infty^2}
    + \left( \frac{3\kappa^2}{8{\bf V} \omega} \right)^2 \,
    e^{-2\omega t} \vev{p_\infty^2} + \frac{3\kappa^2}{8{\bf V}\omega}\left(
    2\Re(\vev{u_\infty p_\infty})\right) \right] .
\label{vevoft}
\ee

It is obvious from Eq.\ref{vevoft} that at large times the volume
behaves as a single exponential, as expected from the solution of
the classical Einstein equations. More interesting, however, is
the fact that letting $t \rightarrow t+t_0$, where $t_0$ is
defined by the condition
\be
    e^{2\omega t_0} = {3\kappa^2 \over 8 {\bf V} \omega}
    \sqrt{\vev{p_\infty^2} \over \vev{u_\infty^2}} ,
\ee
allows us to rewrite Eq.\ref{vevoft} as
\ba
    \vev{V(t)}&=&\frac{3\kappa^2 \sqrt{\vev{u_\infty^2} \vev{p_\infty^2}}}{
    8 \omega} \left[ \cosh(\omega t)
    + \frac{\Re(\vev{u_\infty p_\infty})}{\sqrt{\vev{u_\infty^2} \vev{p_\infty^2}}}
    \right] \nonumber\\
    &=& \frac{\kappa^2\sqrt{\vev{u_\infty^2} \vev{p_\infty^2}}}{4 {\cal H}}
    \left[ \cosh(\omega t)
    + \frac{\Re(\vev{u_\infty p_\infty})}{\sqrt{\vev{u_\infty^2} \vev{p_\infty^2}}}
    \right]
\label{vevtwo}
\ea
Thus, we see $\vev{V(t)}$ corresponds to a system which
is contracting at large times in the past and which then bounces and begins
to re-expand in the future.  During most of this history the system
satisfies the Friedmann equation to high accuracy and expands
(or contracts) with a Hubble constant equal to
\be
    {\cal H} = \frac{2}{3} \omega = \sqrt{\frac{\kappa^2
\Lambda}{3}} .
\ee
However, there is a period in time where the
quantum corrections to the Friedmann equation dominate the
behavior; namely, at times $t \approx 1/\omega$.  Assuming, for
the sake of argument, that were to set $1/\kappa {\cal H} \approx
10^3$, as it is in many models of slow roll inflation, and
assuming $\sqrt{\vev{u_\infty^2} \vev{p_\infty^2}}$ to be of order
unity, then the minimum volume of the universe at the time of the
bounce is on the order of $10^3$ Planck volumes; i.e., on the
order to $10$ Planck-lengths in each dimension.  This sets the
order of magnitude of the scale at which the quantum corrections
become important.   It is gratifying that these quantum
corrections keep the system from contracting forever and ending in
a {\it big crunch\/}.

Another very interesting feature of Eq.\ref{vevtwo} is that it is
characterized by only two numbers, $\sqrt{\vev{u_\infty^2} \vev{p_\infty^2}}$ and
$\Re{\vev{u_\infty p_\infty}}/\sqrt{\vev{u_\infty^2} \vev{p_\infty^2}}$.
The first number is unrestricted in magnitude and roughly determines
the physical volume of the universe at the time of the bounce.
The second number, is constrained by the Schwarz inequality
to lie between plus and minus one, and parameterizes the degree to which
the behavior of the system during the time of the bounce deviates from
a pure hyperbolic cosine.  If the time over which the deviation takes place is
characterized by $1/\omega \approx 1/{\cal H}$, then the minimum size to which
the system contracts is characterized by the ratio of the energy density in the
state to the cosmological constant.  This statement follows from taking the
expectation value of the Hamiltonian as written in Eq.\ref{hamiltonian}, which
implies
\be
    \Re(\vev{u_\infty p_\infty}) = \frac{2}{\kappa \sqrt{3\Lambda}} \vev{{\bf H}} .
\ee
Note, it would appear from the Schwarz inequality that in principle one
could have an observer for whom the universe actually shrinks to zero
size before it bounces.  Fortunately it is easy to see that this can only
occur if $u_0$ or $p_0$ diverges, which violates the condition on allowable
physical states, since such states would have infinite values for
$\vev{{\bf H}^2}$.

Finally, Eq.\ref{vevtwo} shows that any two quantum observers which give the
same values for  $\sqrt{\vev{u_\infty^2} \vev{p_\infty^2}}$ and
$\Re(\vev{u_\infty p_\infty})\sqrt{\vev{u_\infty^2} \vev{p_\infty^2}}$,
see the same physics.  They only differ by the time they see the bounce
occur.  For Gaussian packets we see that this will be true for observers which
are related by the transformation
\be
u_0 \rightarrow \lambda u_0, \,
p_0 \rightarrow \frac{p_0}{\lambda}, \, {\rm and \ }
\gamma \rightarrow \lambda^2 \gamma .
\ee
It is easy to check that this can be implemented by
a unitary transformation.  The values of $u_0$ and $p_0$ can be changed
by means of the shift operators used to define the coherent states in the
first place.  The width of the Gaussian can be changed by application of
a unitary {\it squeezing operator\/} of the form
\be
    e^{\left(\alpha(\gamma)\, {a^\dag}^2 - \alpha(\gamma)^\ast\, a^2\right)} ,
\ee
where the creation and annihilation operators are defined such that
\be
    u_\infty = \frac{1}{\sqrt{2\gamma}}\left(a^\dag + a\right) \quad {\rm and } \quad p_\infty = -i \,
    \sqrt{\frac{\gamma}{2}} \left(a^\dag - a\right) .
\ee

\section{Remarks Concerning The Computation of CMB Anisotropy}

While we have not yet done any detailed computations, it is clear that
the fact that the quantum system deviates from pure exponential growth
at a finite time in the past could have implications for the usual
derivation of CMB fluctuations.  It is entirely possible that the
delay in the time at which the long wavelength modes of the scalar
field exit the horizon relative to the shorter wavelength modes
might produce visible effects in the predicted measurement of
$\delta \rho/\rho$.  If this is so then one should be able to
put an experimental limit on how far back in time one can
push the start of the usual computation.

\section{Minkowski Space  $\Lambda=0$}

Finally, we would like to discuss what happens when we
take $\Lambda = 0$, because, in this case, things work quite a bit differently.
The $\Lambda=0$ Hamiltonian is
\be
    {\bf H} = -\,\frac{3\kappa^2}{16{\bf V}} p_u^2
\ee
and the Heisenberg equations of motion take the form
\be
    {d u \over dt}(t) = -\,\frac{3\kappa^2}{16{\bf V}} p_u  \quad ; \quad
    {d p_u \over dt}(t) = 0 .
\ee
The exact solution to these equations is
\be
    u(t)= u - \A\,p_u\,t   \quad ; \quad  p_u(t) = p_u
\label{heistwo}
\ee
Taking the square of $u(t)$ we obtain the volume operator
\be
    V(t) = {\bf V} u^2(t) = {\bf V}\left[ u^2 - \A\,\left(u\,p_u
    + p_u\,u\right)\,t + \left(\A\right)^2\,p_u^2\,t^2
\right] .
\ee
It follows once again that, as in the de Sitter case, the volume
operator becomes classical at large times in the past and the future.
In this case however there is a state which, while non-normalizable, satisfies the
condition $G(t) \ket{\Psi} = 0$ for all times; namely, the eigenstate of
$p_u$ with eigenvalue $0$.  Now, however, this condition is consistent with
the Heisenberg equations of motion, because in this eigenstate $u(t)=u$ and is
independent of time.  Moreover, this state satisfies
the requirement that $\vev{{\bf H}^2}$ is finite.
Obviously, this state is the limit of sequence
Gaussian packets in $p_u$ of smaller and smaller width.
If we choose this quantum observer then, after we absorb the scale factor into
$\vec{x}$, we find that this observer sees a time-independent Minkowski space.

It is interesting to ask what other, less special, observers see.
Let us assume we are working with an arbitrary coherent state of the form
discussed in the previous section.  Then, the expectation value
of the volume operator is
\be
    \vev{V(t)} = {\bf V} \left[\vev{u^2} - 2\A\,\Re(\vev{u\,p_u})\,t
    + \left(\A\right)^2\vev{p_u^2}\,t^2 \right] ,
\ee
which can be rewritten in the form
\be
\vev{V(t)} = {\bf V} \left[\frac{\vev{u^2}\vev{p_u^2}-\Re(\vev{u\,p_u})^2}{\vev{p_u^2}}
+\left(\frac{3\kappa^2}{16 {\bf V}}\right)^2 \,\vev{p_u^2}
\left(t - \frac{16{\bf V}\Re(\vev{u\,p_u})}{3\kappa^2\,\vev{p_u^2}}
\right)^2\right] .
\ee
Thus, we see that for the generic observer, the case of zero cosmological
constant actually corresponds to a universe for which the volume factor
is expanding like $t^2$, or for which the scale factor $a(t)$ is growing
like $t^{2/3}$.  Surprisingly this corresponds to a universe dominated by
non-relativistic matter.  In other words, a non-vanishing energy
density present in the quantum excitations of the scale factor
produce the same effect as cold matter.

A final point worth mentioning is that, as in the case of de Sitter space,
the Schwarz inequality guarantees that the volume never shrinks to
zero for any allowable physical observer; i.e., we never are in the situation of
a {\it big crunch}.  It is interesting to note that in this formalism the big
crunch is averted due to the quantum physics of the long wavelength modes of
the gravitational field and not short distance physics.

\section{Summary}

In our first paper we outlined a general formalism for setting up a fully quantum
calculation of the CMB fluctuations, including back reaction.  We also
suggested a pixelization scheme which should allow us to extend this computation
to include non-linear quantum effects for a finite number of long-wavelength modes
of both the Newtonian potential and matter fields.  In this paper we applied the
general formalism to the case of de Sitter and Minkowski space in order to show
in an explicit, exactly solvable, case how the formalism works in detail and why
we are generically forced to choose to impose the classical constraint condition
as an asymptotic condition on allowable quantum states.  The most important
result of our discussion is that in the case of de Sitter space the
system deviates from the expected pure exponential expansion at a finite
time in the past.  One possible consequence of this fact, is that one might be able to
experimentally measure the effects of the quantum corrections to the pure Einstein
equations as deviations from the conventionally predicted form of
$\delta \rho/\rho$.  Failing
that, one might be able to bound the earliest time at which one is free to
set initial conditions on the state of the inflaton and other fields
in the system\cite{Guth:prd23}.  In other words,
either there may well be measureable consequences following from
the quantum nature of the problem at early times, or one will have to face
up to the problem of how and when to set initial conditions.

While, as it stands, the formalism we have presented is by no means a good
candidate for a theory of everything, we feel that the interesting results
obtained by proceeding along these lines suggests it is a very good
candidate for a theory of something.  Namely, a fully quantum theory
of the measured fluctuations in the CMB radiation.

\section{Acknowledgements}

We would like to thank J.~D.~Bjorken for helpful communications.

\end{document}